\documentclass{SciPost}

\hypersetup{
    colorlinks,
    linkcolor={red!50!black},
    citecolor={blue!50!black},
    urlcolor={blue!80!black}
}

\usepackage[bitstream-charter]{mathdesign}
\usepackage[capitalise]{cleveref}
\usepackage{listings}
\usepackage{float}
\usepackage{physics}
\usepackage{braket}
\usepackage{microtype}
\usepackage{inconsolata}
\lstdefinelanguage{Julia}%
  {morekeywords={abstract,break,case,catch,const,continue,do,else,elseif,%
      end,export,false,for,function,immutable,import,importall,if,in,%
      macro,module,mutable,struct,otherwise,quote,return,switch,true,try,type,typealias,%
      using,while},%
   sensitive=true,%
   alsoother={$},%
   morecomment=[l]\#,%
   morecomment=[n]{\#=}{=\#},%
   morestring=[s]{"}{"},%
   morestring=[m]{'}{'},%
}[keywords,comments,strings]%

\newfloat{lstfloat}{htbp}{lop}
\floatname{lstfloat}{Listing}

\crefalias{lstfloat}{listing}

\DeclareSymbolFont{usualmathcal}{OMS}{cmsy}{m}{n}
\DeclareSymbolFontAlphabet{\mathcal}{usualmathcal}

\fancypagestyle{SPstyle}{
\fancyhf{}
\lhead{\colorbox{scipostblue}{\bf \color{white} ~SciPost Physics Codebases }}
\rhead{{\bf \color{scipostdeepblue} ~Submission }}

\fancyfoot[C]{\textbf{\thepage}}
}

\newcommand\banivo{BaNi\textsubscript{2}V\textsubscript{2}O\textsubscript{8}}

\begin{document}

\pagestyle{SPstyle}

\begin{center}{\Large \textbf{\color{scipostdeepblue}{
Carlo.jl: A general framework for Monte Carlo simulations in Julia
}}}\end{center}

\begin{center}\textbf{
Lukas Weber\textsuperscript{1,2$\star$}
}\end{center}

\begin{center}
{\bf 1} Center for Computational Quantum Physics, The Flatiron Institute,
162 Fifth Avenue, New York, New York 10010, USA
\\
{\bf 2} Max Planck Institute for the Structure and Dynamics of Matter,
Luruper Chaussee 149, 22761 Hamburg, Germany
\\[\baselineskip]
$\star$ \href{mailto:lweber@flatironinstitute.com}{\small lweber@flatironinstitute.com}
\end{center}

\section*{\color{scipostdeepblue}{Abstract}}
{\boldmath\textbf{%
Carlo.jl is a Monte Carlo simulation framework written in Julia. It provides MPI-parallel scheduling, organized storage of input, checkpoint, and output files, as well as statistical postprocessing. With a minimalist design, it aims to aid the development of high-quality Monte Carlo codes, especially for demanding applications in condensed matter and statistical physics. This hands-on user guide shows how to implement a simple code with Carlo.jl and provides benchmarks to show its efficacy.
}}

\vspace{\baselineskip}

\noindent\textcolor{white!90!black}{%
\fbox{\parbox{0.975\linewidth}{%
\textcolor{white!40!black}{\begin{tabular}{lr}%
  \begin{minipage}{0.6\textwidth}%
    {\small Copyright attribution to authors. \newline
    This work is a submission to SciPost Physics Codebases. \newline
    License information to appear upon publication. \newline
    Publication information to appear upon publication.}
  \end{minipage} & \begin{minipage}{0.4\textwidth}
    {\small Received Date \newline Accepted Date \newline Published Date}%
  \end{minipage}
\end{tabular}}
}}
}

\vspace{10pt}
\noindent\rule{\textwidth}{1pt}
\tableofcontents
\noindent\rule{\textwidth}{1pt}
\vspace{10pt}

\section{Introduction}
\label{sec:intro}
Monte Carlo methods have been a powerful computational tool since the very beginnings of the fields of statistical and condensed matter physics\cite{Metropolis1953}. Initially built to study the classical Boltzmann distribution, they perform well on a wide variety of high-dimensional integration problems due to their ability to break the curse of dimensionality. In quantum systems, where a probabilistic interpretation of the partition function is in general not possible, quantum Monte Carlo algorithms can exploit certain quantum-to-classical mappings to make powerful and accurate predictions for models that are out of reach of other state-of-the art methods~\cite{Gubernatis2016}. 

In exchange for simulating high-dimensional problems, quantum Monte Carlo simulations in particular rely on efficient implementations making use of massive parallelism. In the course of a simulation, many random samples are produced that have to be stored and statistically postprocessed to arrive at results with reliable errorbars. These requirements typically necessitate some amount of ``bookkeeping'' code, unrelated to the Monte Carlo algorithm itself, but still crucial to its performance. This inspired the development of earlier frameworks such as the ones included in ALPS~\cite{ALPS} and the subsequent ALPSCore~\cite{ALPSCore} in C++.

The Julia programming language~\cite{Bezanson2012} is becoming increasingly popular as an alternative to traditional scientific computing languages such as C++ and Fortran. Its heavy reliance on type inference, dynamic dispatch, and just-in-time compilation allow it to provide both high performance and the flexibility of a high-level scripting language. Furthermore, its modern approach to dependency management greatly facilitates installing and reusing existing software.

The purpose of Carlo.jl is to help implementers of Monte Carlo codes in Julia to write high-quality, high-performance packages without worrying about the peripheral bookkeeping tasks that are common to all Monte Carlo simulations. Its features include (i) Monte Carlo aware parallel scheduling (including parallel tempering) (ii) organized and reproducible storage of simulation inputs, checkpoints, and results, as well as (iii) statistical postprocessing of autocorrelated Monte Carlo samples.

In \cref{sec:features}, the features of Carlo.jl and their inner workings are explained in detail. \cref{sec:examples} contains three examples: an Ising code that provides a minimal skeleton for implementing Monte Carlo algorithms in Carlo.jl, a state-of-the-art stochastic series expansion Monte Carlo code~\cite{Sandvik1999}, and an example of the parallel tempering feature. Afterwards, in \cref{sec:benchmarks}, we show correctness and performance benchmarks of Carlo.jl, including a direct performance comparison of a Julia and a C++ implementation of the stochastic series expansion code. Finally, in \cref{sec:conclusion}, a short conclusion and outlook are given.

Apart from this user guide, users may refer to the Carlo.jl documentation\footnote{The documentation can be viewed online on \href{https://github.com/lukas-weber/Carlo.jl.git}{GitHub}, built locally by running  \texttt{cd docs; julia ----project make.jl}, or accessed directly within the Julia REPL.} for more detailed descriptions of each component.

\section{Features}
\label{sec:features}
The purpose of this section is to give a high-level overview of the way Carlo.jl works without descending too far into technical details and concrete usage instructions. For a more hands-on introduction into how to use Carlo.jl in practice, see \cref{sec:examples}.
\subsection{Monte Carlo aware parallel scheduling}
One useful property of Monte Carlo simulations is that they are embarrassingly parallel. An arbitrary number of random processes can be allocated to many workers in parallel, which, once thermalized, can all produce samples that are finally averaged together. Apart from this intrinsic parallelism, simulations are often run for a variety of different parameters in parallel.
\begin{figure}
\includegraphics{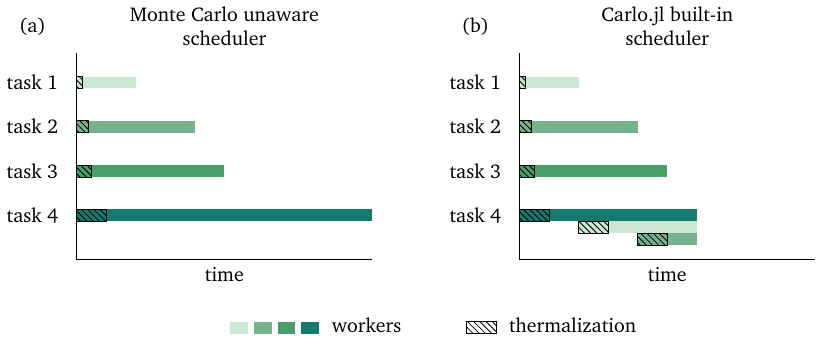}
\caption{\textbf{Parallelism in Monte Carlo simulations}. (a) Monte Carlo unaware parallelization over different parameters (e.g. system sizes, temperatures) can lead to suboptimal resource usage if tasks are not well balanced. (b) Carlo.jl parallelizes over different parameter sets but also takes advantage of the possibility to share work between different Monte Carlo processes for the same task. Each new process needs to thermalize first, which is a serial cost that can only be reduced by more fine-grained parallelization.}
\label{fig:mcsched}
\end{figure}
Both of these kinds of parallelism can be exploited using a standard parallel scheduler. However, if the load between different parameter sets is not balanced, the resource usage of this approach is not optimal: Most workers can run out of work while few are still stuck simulating the last few expensive parameter sets (\cref{fig:mcsched}(a)). In Monte Carlo simulations, such a situation can always be avoided by simply allowing the idle workers to join other tasks and share the work (\cref{fig:mcsched}(b)).

Carlo.jl uses MPI.jl~\cite{Byrne2021} to orchestrate this kind of Monte Carlo aware, work-sharing scheduling and automatically averages the results. Each parameter set in the simulation corresponds to a \textit{task} and each task may have multiple \textit{runs}, which correspond to independent random processes that are averaged. If workers are idle, the scheduler will assign them a new task from the set of unfinished tasks in a round-robin manner and start a new run for that task, provided the remaining work is not less that the number of thermalization sweeps.

On top of this model of trivial parallelism, Carlo.jl also supports nontrivial MPI parallelism: The work of each single run may optionally be shared by a team of MPI ranks. This advanced feature, called \textit{parallel run mode} allows the natural implementation of certain Monte Carlo meta-techniques such as parallel tempering, which will be illustrated in \cref{sec:parallel_tempering}. Another possible use case is the implementation of branching-random walk codes that propagate an population of walkers in parallel.
\subsection{Organized storage of inputs, checkpoints and results}
As mentioned in the previous section, large-scale Monte Carlo simulations can involve scanning large parameter spaces and produce large numbers of random samples. Long-running simulations should also implement some form of checkpointing so that valuable computation time is not wasted when an error occurs or the job runs out of its allocated runtime on a computing cluster. 

These boundary conditions induce data management challenges at the beginning, during, and at the end of a simulation. This section explains how Carlo.jl handles these challenges for the user.

\subsubsection{Input}
Rather than specifying simulation parameters in traditional configuration files, Carlo.jl leverages the dynamic nature of Julia to set parameters programmatically. The recommended workflow for this is to write a job script as shown in \cref{code:input}. The job script can be split into three parts. First, the \texttt{TaskMaker} from \texttt{Carlo.JobTools} is used to procedurally generate a list of parameter sets. Then, this list is retrieved using \texttt{make\_{}tasks(tm)} and fed into the \texttt{JobInfo} alongside other options such as the maximum runtime limit and the checkpoint interval. Finally, \texttt{start(job, ARGS)} initiates the Carlo.jl command line interface (CLI) when the job script is executed.
\begin{lstfloat}
\begin{lstlisting}[language=julia]
# example_job.jl

using Carlo
using Carlo.JobTools
using Ising

tm = TaskMaker()

tm.sweeps = 20000
tm.thermalization = 2000
tm.binsize = 100

Ts = range(1, 3, 10)
Ls = [8, 12, 16]
for L in Ls
    for T in Ts
        tm.T = T
        tm.Lx = L
        tm.Ly = L
        task(tm)
    end
end

job = JobInfo("example_job", Ising.MC;
    run_time = "24:00:00",
    checkpoint_time = "30:00",
    tasks = make_tasks(tm),
)
start(job, ARGS)
\end{lstlisting}
\caption{\textbf{Carlo job script}. This snippet generates a job for the Ising code implemented in \cref{sec:ising}. Arbitrary parameters can be assigned to the \texttt{TaskMaker} \texttt{tm}. The function \texttt{task(tm)} creates a snapshot of all currently set values and turns it into a task that will be simulated. The list of tasks is returned by \texttt{make\_{}tasks} and passed to the \texttt{JobInfo} structure. The detailed meanings of the options are explained in \cref{sec:ising}.}
\label{code:input}
\end{lstfloat}
The CLI allows \textit{running} the simulation, and for unfinished simulations allows checking their \textit{status} and \textit{merging} the already available data to obtain preliminary results. 

While recommended, the use of \texttt{TaskMaker} and the CLI is optional. For more information on how to manually start simulations from within Julia, interested readers are referred to the documentations of \texttt{JobInfo} and the \texttt{start} function.
\subsubsection{Checkpointing}
In a high-performance computing environment, calculations often run on limited time allocations that may be shorter than the length of the calculation one wishes to perform. If a calculation runs out of time or uses too much memory, it may be terminated by the environment, which leads to the loss the entire progress.

Checkpointing, i.e. saving the complete Monte Carlo configuration to disk at regular intervals, allows to gracefully recover from such situations and simply continue the calculation from the point it left off. Carlo provides a simple interface to save the configuration state to an HDF5~\cite{HDF5, HDF5jl} group. Internally, it makes sure that these checkpoints are never corrupted, even if the process is terminated in the middle of writing.\footnote{This works by writing to a temporary file first and then moving that file to the final location, overwriting the old file. Moving (as opposed to writing) is an atomic operation that cannot be interrupted.} Running a simulation as described in the previous subsection will automatically resume from checkpoints if they exist.

While packages like JLD2.jl~\cite{JLD2} exist that allow automatically saving arbitrary Julia objects, the Carlo.jl checkpointing interface instead relies on manually choosing which data to save in what way. This can be helpful in reducing the checkpoint file size and encourages maintaining stable serialization interfaces.
\subsubsection{Results}
Like checkpoints, Carlo.jl stores simulation results in HDF5 files. Inspecting these files allows retrieving the entire Monte Carlo time series. In the postprocessing step that is described in \cref{sec:postproc}, these raw files are merged and averaged to produce means, statistically correct error bars and approximate autocorrelation times. These merged results are saved in a human-readable JSON file. The \texttt{Carlo.ResultTools} module provides a simple way to import these JSON files into a Julia Dataframe (e.g. from \texttt{DataFrames.jl}\cite{BouchetValat2023}).

To facilitate reproducibility, and conform with the best practices of research data management, the result files produced by Carlo.jl contain all parameters used to run the simulation as well as the version of Carlo.jl and of the package containing the Monte Carlo algorithm implementation.

\subsection{Postprocessing}
\label{sec:postproc}
Monte Carlo time-series are in general autocorrelated. Carlo.jl performs postprocessing on measured observables to provide the user with statistically sound means and standard errors using a binning analysis.

The binning analysis works in two steps. First, an \textit{internal binning} step is performed during the simulation before Monte Carlo measurements are written to disk. The bin size for this step is set by \texttt{binsize} in the task parameters. The purpose of internal binning is simply to reduce memory and disk-I/O overhead. As long as \texttt{binsize} is not big compared to the total number of samples, it does not influence the magnitude of the final errors computed by Carlo.jl.

Once the simulation is complete, for a given task, there will be multiple files containing measurement results from different independent runs that were scheduled by Carlo.jl (\cref{fig:mcsched}(b)). In the second step of the binning analysis, these internally binned Monte Carlo time series are concatenated and binned further. The purpose of this ``rebinning'' step is to produce sufficiently large bins to recover uncorrelated statistics that allow extracting the statistical errors using the standard estimator

\begin{equation}
\sigma_A = \sqrt{\frac{1}{N_\text{rebin}(N_\text{rebin}-1)} \sum_n \qty(\overline{A}_n-\overline{A})^2},
\end{equation}
where $\overline{A}_n$ are the means of each rebinned sample and $\overline{A} = \sum_n \overline{A}_n / N_\text{rebin}$.
By default, the number of “rebins” $N_\text{rebin}$ is chosen using the heuristic
\begin{equation}
N_\text{rebin}(N) = N_\text{rebin}^\text{min} + \max\{0, \sqrt[3]{N-N_\text{rebin}^\text{min}}\}
\end{equation}
where $N$ is the total number of measured bins and $N_\text{rebin}^\text{min}=10$ is a minimum in case there are only few bins in the simulation. Choosing a cube-root scaling is a simple way to ensure that both $N_\text{rebin}$ and the length of each rebin, $N/N_\text{rebin}$, approach infinity for increasing $N$, while devoting more resources to the latter. The default value of $N_\text{rebin}$ can be overriden using the task parameter \texttt{rebin\_{}length} in the job script.

A rough estimate for the autocorrelation time $\tau_\text{auto}$ is calculated based on the ratio of the rebinned error $\sigma_A$ and naive error computed without any rebinning, $\tilde{\sigma}_A$,~\cite{Ambegaokar2010}
\begin{equation}
\label{eq:autocorr}
\tau_\text{auto} = \frac{\sigma_A^2}{2\tilde{\sigma}^2_A}.
\end{equation}
Note that $\tau_\text{auto}$ is measured in units of the internal bin size (set by the \texttt{binsize} parameter). To find the autocorrelation time in terms of Monte Carlo sweeps, \texttt{binsize} should be set to 1.

 Apart from the observables themselves, users can also define \textit{evaluables} that depend on the observables via some function. The bias-corrected means and errors for evaluables are then 
propagated using the jackknife~\cite{Miller1974} method on the rebinned samples.
\section{Examples}
\label{sec:examples}
The specifics of implementing and running codes with Carlo.jl are best explained using examples. In the following, we will show three examples, in raising complexity.

The first example, \cref{sec:ising}, will be the paradigmatic Metropolis Monte Carlo algorithm for the Ising model. While this algorithm itself is very simple and of limited use in practice, it provides a minimal skeleton for implementing new Monte Carlo algorithms in Carlo.jl without the distractions of method-specific details.

The second example, \cref{sec:sse}, will be an implementation of the stochastic series expansion Monte Carlo~\cite{Sandvik1999} approach on top of Carlo.jl. While this implementation is a separate package and we will not go into details about its inner workings, it here serves to show how Carlo.jl is useful in state-of-the-art Monte Carlo calculations. This will be supplemented in \cref{sec:benchmarks} by a benchmark directly comparing Julia and C++ performance for that code.

The last example, \cref{sec:parallel_tempering}, will have a split purpose. First, it will introduce the parallel-tempering feature that is included in Carlo.jl. Second, it will show how parallel-tempering was implemented as a simple wrapper on top of the parallel run mode feature.
\subsection{Ising model}
\label{sec:ising}
In this section, we will give a hands-on usage example for Carlo.jl by implementing and running the Metropolis Monte Carlo algorithm~\cite{Metropolis1953} for the Ising model on a square lattice,
\begin{equation}
H = -\sum_{x,y} \sigma^z_{x,y} (\sigma^z_{x+1,y} + \sigma^z_{x,y+1}),
\end{equation}
with periodic boundary conditions.
All code for this example is available in a separate repository~\cite{IsingExample}.

\subsubsection{Implementation}
First, we need to implement the algorithm. It is a good idea to write such an implementation as its own Julia package. In the following, we will assume that we have generated a package called \texttt{Ising}, with a file \texttt{Ising.jl} containing all the code.

In this file, we declare a new subtype of \texttt{AbstractMC}.
\begin{lstlisting}[language=julia]
# Ising.jl
module Ising

using Carlo
using HDF5

struct MC <: AbstractMC
    T::Float64

    spins::Matrix{Int8}
end
\end{lstlisting}
The type \texttt{MC} holds two fields: the temperature \texttt{T} and a matrix of integers representing the spins on the square lattice.

The constructor of \texttt{MC} should accept a dictionary of parameters that are defined in the jobscript (see \cref{code:input}).
\begin{lstlisting}[language=julia]
function MC(params::AbstractDict)
    Lx = params[:Lx]
    Ly = get(params, :Ly, Lx)
    T = params[:T]
    return MC(T, zeros(Lx, Ly))
end
\end{lstlisting}
We read the temperature and lattice dimensions from the parameters and construct the \texttt{MC} structure accordingly.

The remainder of the implementation consists of methods for Carlo.jl's \texttt{AbstractMC} interface.

\begin{lstlisting}[language=julia]
function Carlo.init!(mc::MC, ctx::MCContext, params::AbstractDict)
    mc.spins .= rand(ctx.rng, Bool, size(mc.spins)) .* 2 .- 1
    return nothing
end
\end{lstlisting}
The \texttt{Carlo.init!} function is called to initialize the Monte Carlo configuration at the beginning of a new simulation (but not when restarting from a checkpoint).

In addition to the \texttt{MC} object, it receives a \texttt{MCContext}, which is our handle to interact with Carlo.jl's functionality. Here, we use the integrated random number generator \texttt{ctx.rng} to generate a spin configuration. It is important to always use \texttt{ctx.rng} for random numbers to maintain reproducibility from a given random seed.\footnote{The seed can be set in the jobscript via the parameter \texttt{seed}.}

Next we implement the Monte Carlo updates or sweeps.
\begin{lstlisting}[language=julia]
function periodic_elem(spins::AbstractArray, x::Integer, y::Integer)
    return spins[mod1.((x, y), size(spins))...]
end

function Carlo.sweep!(mc::MC, ctx::MCContext)
    Lx = size(mc.spins, 1)

    for _ = 1:length(mc.spins)
        i = rand(ctx.rng, eachindex(mc.spins))
        x, y = fldmod1(i, size(mc.spins,1))

        neighbor(dx, dy) = periodic_elem(mc.spins, x + dx, y + dy)
        ratio = exp(
            -2.0 / mc.T *
            mc.spins[x, y] *
            (neighbor(1, 0) + neighbor(-1, 0)
            + neighbor(0, 1) + neighbor(0, -1)),
        )

        if ratio >= 1 || ratio > rand(ctx.rng)
            mc.spins[x, y] *= -1
        end
    end
    return nothing
end
\end{lstlisting}
According to the Metropolis algorithm, we propose a spin flip and calculate the energy difference between the new and old configurations, which can be reduced to a calculation of the neighboring spins. Note the use of functions like \texttt{mod1} and \texttt{fldmod1}, which are helpful when applying modular arithmetic to the 1-based indices of Julia arrays.

Next are the measurements.
\begin{lstlisting}[language=julia]
function Carlo.measure!(mc::MC, ctx::MCContext)
    mag = sum(mc.spins) / length(mc.spins)

    energy = 0.0
    correlation = zeros(size(mc.spins, 1))

    for y = 1:size(mc.spins, 2)
        for x = 1:size(mc.spins, 1)
            neighbor(dx, dy) = periodic_elem(mc.spins, x + dx, y + dy)
            energy += -mc.spins[x, y]*(neighbor(1,0)+neighbor(0,1))

            # in practice, one should use more lattice symmetries!
            correlation[x] += mc.spins[1, y] * mc.spins[x, y]
        end
    end

    measure!(ctx, :Energy, energy / length(mc.spins))

    measure!(ctx, :Magnetization, mag)
    measure!(ctx, :AbsMagnetization, abs(mag))
    measure!(ctx, :Magnetization2, mag^2)
    measure!(ctx, :Magnetization4, mag^4)

    measure!(ctx, :SpinCorrelation, correlation ./ size(mc.spins, 2))
    return nothing
end
\end{lstlisting}
We calculate the magnetization and its moments, the energy per spin, and the spin correlation function. The function \texttt{measure!(ctx, name, value)} passes each result to Carlo.jl. As we see for the correlation function, observables need not be scalars. Vectors (or higher-rank) arrays are also supported.

The next method is optional, but can be used to register \textit{evaluables}, that is, quantities that depend on Monte Carlo expectation values. Carlo.jl will perform error-propagation and bias correction on these (see~\cref{sec:postproc}).
\begin{lstlisting}[language=julia]
function Carlo.register_evaluables(
	::Type{MC}, eval::AbstractEvaluator, params::AbstractDict
)
    T = params[:T]
    Lx = params[:Lx]
    Ly = get(params, :Ly, Lx)

    evaluate!(eval, :BinderRatio,
        (:Magnetization2, :Magnetization4)
    ) do mag2, mag4
        return mag2 * mag2 / mag4
    end

    evaluate!(eval, :Susceptibility, (:Magnetization2,)) do mag2
        return Lx * Ly * mag2 / T
    end

    evaluate!(eval, :SpinCorrelationK, (:SpinCorrelation,)) do corr
        corrk = zero(corr)
        for i = 1:length(corr), j = 1:length(corr)
            corrk[i] += corr[j] * cos(2pi/length(corr)*(i-1)*(j-1))
        end
        return corrk
    end

    return nothing
end
\end{lstlisting}
Using \texttt{evaluate!}, we provide formulas for useful quantities such as the Binder ratio, the susceptibility and the Fourier transform of the correlation function. Note that \texttt{Carlo.register\_{}evaluables} is called during the postprocessing step, after the simulation is complete. It has no access to any simulation state.

Finally, we implement the checkpointing interface.
\begin{lstlisting}[language=julia]
function Carlo.write_checkpoint(mc::MC, out::HDF5.Group)
    out["spins"] = mc.spins
    return nothing
end

function Carlo.read_checkpoint!(mc::MC, in::HDF5.Group)
    mc.spins .= read(in, "spins")
    return nothing
end

end # module Ising
\end{lstlisting}
When starting a simulation from a checkpoint, \texttt{Carlo.read\_{}checkpoint!} is called instead of \texttt{Carlo.init!}, so that we have to recover everything from disk that is not already calculated in the constructor \texttt{MC(params)}. The temperature is set in the constructor so we do not have to save it to disk. The spin configuration, however, has to be saved. For more information on how to read and write to an \texttt{HDF5.Group}, see the documentation of HDF5.jl~\cite{HDF5}.

This concludes the \texttt{Ising.jl} file and our implementation. Next, we will use it to run some simulations.

\subsubsection{Running simulations}
\label{sec:runising}
To run a simulation, we first need to install Carlo.jl and the Ising module. This works via the Julia REPL.
\begin{lstlisting}
julia>]
pkg> add Carlo
pkg> add https://github.com/lukas-weber/Ising.jl
\end{lstlisting}
Readers who have followed along in the last section and wrote their own code in a different directory, can instead import it in development mode using \texttt{dev path/to/Ising}.

Next, we write the job script from \cref{code:input}. The meanings of the parameters are as follows.
\begin{description}
\item[\texttt{Lx}, \texttt{Ly}, \texttt{T}] used by our implementation to set the size and temperature of the Ising model.
\item[\texttt{thermalization}] the number of sweeps that are performed in the thermalization phase of the simulation, before measurements are taken.
\item[\texttt{sweeps}] the number of measurement sweeps.
\item[\texttt{binsize}] the internal bin size, that is the number of samples that are averaged before writing to disk. This should be much smaller than the total number of sweeps but large enough to not waste too much disk space in large simulations.
\end{description}
There are more optional parameters that are explained in greater detail in the documentation for \texttt{TaskInfo} and \texttt{JobInfo}.

Using the given system sizes and temperatures, we generate a collection of tasks. Then, all information about the job is fed to the \texttt{JobInfo} constructor, which additionally takes the job name, here \texttt{"example\_{}job"}, the Monte Carlo type we implemented, \texttt{Ising.MC}, as well as the total runtime and the interval between checkpoints.

Finally, \texttt{start(job, ARGS)} invokes the Carlo.jl CLI, which we can run from the shell as follows.
\begin{lstlisting}
$ julia example_job.jl --help
$ julia example_job.jl run
\end{lstlisting}
The first line gives an overview over the CLI, the second line starts the simulation.
For long-running simulations, the commands \texttt{status} and \texttt{merge} are helpful to check the progress and postprocess those results that are already available, respectively.

\begin{figure}
\begin{center}
\includegraphics[scale=0.625]{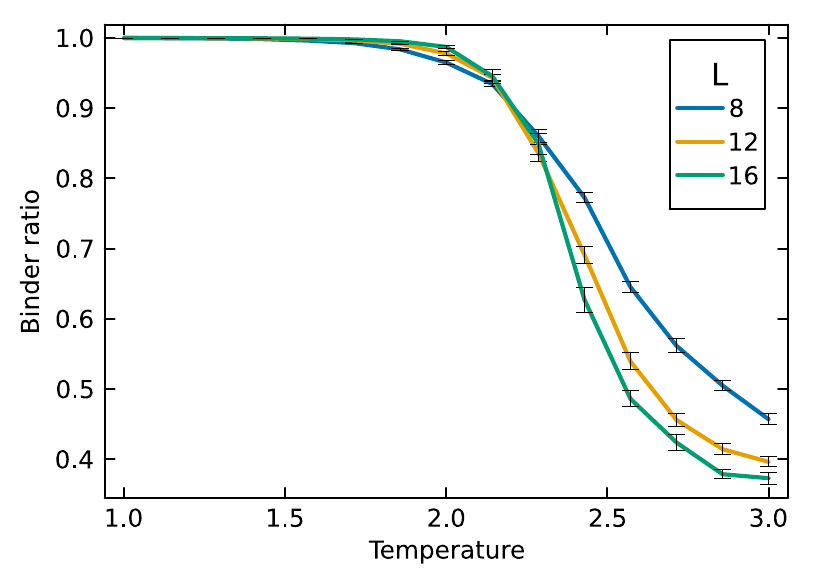}
\end{center}
\caption{\textbf{Binder ratio}, $\braket{m^2}^2/\braket{m^4}$, of the magnetization $m$, in the $L\times L$ square-lattice Ising model as a function of temperature. As one would expect, there is a crossing close to the critical temperature $T_c \approx 2.269$. The error bars are calculated automatically by Carlo.jl using the jackknife method.}
\label{fig:binder}
\end{figure}

After the simulation has completed, we should see a directory \texttt{example\_{}job.data}, which contains raw data and check points, and a file \texttt{example\_{}job.results.json}, which contains the postprocessed results of our simulation.
The submodule \texttt{Carlo.ResultTools} provides tools for reading these results back into Julia.
\begin{lstlisting}[language=julia]
using Plots
using DataFrames
using Carlo.ResultTools

df = DataFrame(ResultTools.dataframe("example_job.results.json"))
plot(df.T, df.BinderRatio;
    xlabel = "Temperature",
    ylabel = "Binder ratio",
    group = df.Lx,
    legendtitle = "L",
)
\end{lstlisting}
The output of this snippet can be seen in \cref{fig:binder}.

To inspect a raw Monte Carlo time series, we can read it directly the respective HDF5 file. For example, for the first temperature and system size, we could do
\begin{lstlisting}[language=julia]
using HDF5

samples = h5read(
    "example_job.data/task0001/run0001.meas.h5",
    "observables/Energy/samples",
)
\end{lstlisting}
to retrieve a vector containing each recorded sample of the energy.

In the example above, we ran Carlo.jl with a single processor. In order to use the parallel scheduler, we can simply launch the Carlo.jl CLI with MPI.
\begin{lstlisting}
$ mpirun -n $ncores julia example_job.jl run
\end{lstlisting}
If the simulation has already been completed before, this will stop without doing further work. To restart the simulation, we should use the flag \texttt{run -r} or call the \texttt{delete} command before running.

In production runs, it is advisable to configure MPI.jl and HDF5.jl to use the system-provided binaries.
\begin{lstlisting}[language=julia]
using MPIPreferences
MPIPreferences.use_system_binary()
using HDF5
HDF5.API.set_libraries!(
    "path/to/libhdf5.so",
    "path/to/libhdf5_hl.so",
)
\end{lstlisting}
Further, by default, Julia will use multithreading for linear algebra operations. Without care, together with MPI parallelization, this can cause overusage of the available CPUs, so 
it should be controlled, e.g. by setting the environment variable \texttt{OPENBLAS\_{}NUM\_{}THREADS=1}.
\subsection{Stochastic series expansion}
\label{sec:sse}
In the last section, we covered the Metropolis algorithm for the Ising magnet. More complicated models of magnetism -- or other quantum phenomena -- generally contain noncommuting operators, which require more sophisticated quantum Monte Carlo algorithms to solve. One quantum Monte Carlo method that is particularly effective for lattice models of bosons is the stochastic series expansion~\cite{Sandvik1999}. In basic terms, it works by expanding the trace over Boltzmann operator $Z=\Tr e^{-\beta H}$ with the Hamiltonian $H$ and the inverse temperature $\beta$ into a power series and performing the resulting sums over over operator products using Monte Carlo sampling. More details can be found, e.g. in Ref.~\cite{Sandvik2010}.

StochasticSeriesExpansion.jl~\cite{WeberStochasticSeriesExpansion2024} is an implementation of the abstract-loop flavor\cite{Weber2022} of the algorithm that can flexibly handle any model consisting of bond operators acting on a graph of bonds (as long as the sign problem~\cite{Pan2024} is not severe). It is built in Julia on top of Carlo.jl, based on the C++ code that was used in earlier publications to simulate frustrated magnets in cluster bases~\cite{Jimenez2021,WeberCluster2022,WeberThermal2022} and spin-photon coupled systems~\cite{weber_cavityrenormalized_2023}.

While this arbitrary model capability requires some code to specify the Hamiltonians and associated observables, some models are supported out of the box. One example, that we will explore in this section are arbitrary-spin anisotropic Heisenberg models of the form
\begin{equation}
\label{eq:heisenberg}
H = \sum_{\braket{i,j}} J_{ij} \qty(\vb S_i \cdot \vb S_j + d_{ij} S^z_i S^z_j) + \sum_i h_i^z S^z_i + D^x_i (S^x_i)^2 + D^z_i (S^z_i)^2,
\end{equation}
where $\braket{i,j}$ sums over the bonds of an arbitrary lattice. As long as the signs of $J_{ij}$ in combination with the bonds of the lattice do not imply magnetic frustration, the model is sign-problem free. This is the case, e.g., for $J_{ij} > 0$ on a bipartite lattice.

As a registered package, we can install StochasticSeriesExpansion.jl (and Carlo.jl if we have not already) from the Julia REPL.
\begin{lstlisting}
julia>]
pkg> add StochasticSeriesExpansion
\end{lstlisting}
Then we can write a job script. In this example, we calculate the susceptibility of an effective model for the anisotropic antiferromagnet \banivo from Ref.~\cite{Klyushina2021}, which is also available as an example in the StochasticSeriesExpansion.jl documentation.
\begin{lstlisting}[language=julia]
# bani2v2o8.jl

using Carlo
using Carlo.JobTools
using StochasticSeriesExpansion

tm = TaskMaker()
tm.sweeps = 80000
tm.thermalization = 10000
tm.binsize = 100

temperatures = range(0.05, 4, 20)
system_sizes = [10, 20]

tm.model = MagnetModel
tm.S = 1
tm.J = 1
tm.Dz = 0.005645 # D^{QMC}_{EP(XY)}/J^{QMC}_n from PRB 104, 065502 

tm.measure = [:magnetization]
for L in system_sizes
    tm.lattice = (unitcell = UnitCells.honeycomb, size = (L, L))

    for T in temperatures
        tm.T = T
        task(tm)
    end
end

job = JobInfo(
    splitext(@__FILE__)[1],
    StochasticSeriesExpansion.MC;
    run_time = "24:00:00",
    checkpoint_time = "30:00",
    tasks = make_tasks(tm),
)

start(job, ARGS)
\end{lstlisting}
The meanings of the StochasticSeriesExpansion.jl-specific parameters are the following
\begin{description}
\item[\texttt{model}] the model to simulate, here \texttt{MagnetModel} which corresponds to \cref{eq:heisenberg}.
\item[\texttt{S}] the spin magnitude
\item[\texttt{T}] the temperature
\item[\texttt{J,d,h,Dx,Dz}] correspond to the parameters in \cref{eq:heisenberg}. Can also be made site-dependent, see StochasticSeriesExpansion.jl documentation.
\item[\texttt{lattice}] the unitcell and the dimensions of the supercell. Honeycomb and other popular ones are predefined. Otherwise, an instance of \texttt{UnitCell} can be passed.
\item[\texttt{measure}] specifies the observables to be measured (on top of the energy and specific heat). Here, just measure uniform magnetization-related quantities. \texttt{:staggered\_{}magnetization} would additionally compute the staggered magnetization. There is an operator string estimator interface for implementing custom observable estimators.
\end{description}
For details, the reader is referred to the StochasticSeriesExpansion.jl documentation.

\begin{figure}
\begin{center}
\includegraphics[scale=0.625]{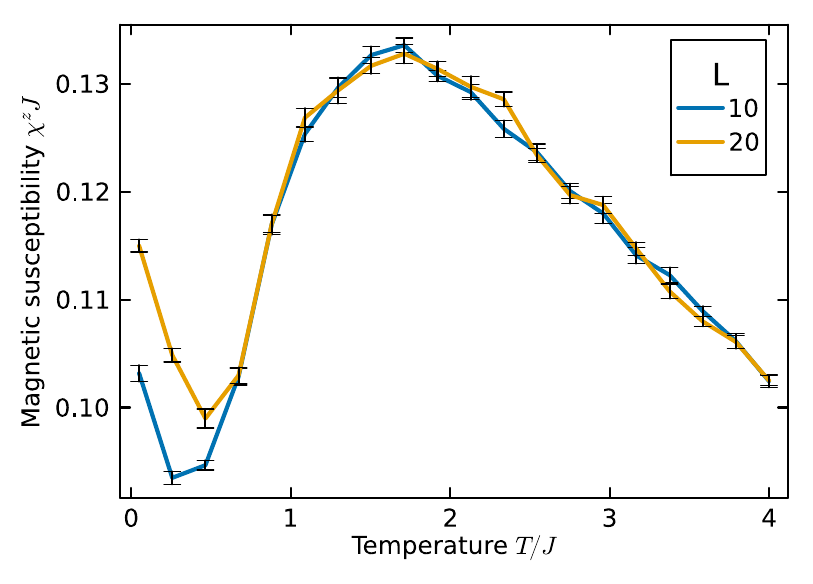}
\end{center}
\caption{\textbf{Example calculation with StochasticSeriesExpansion.jl}. The magnetic susceptibility perpendicular to the easy plane, $\chi^z$, of an effective model for \banivo, as a function of the temperature $T$.}
\label{fig:susceptibility}
\end{figure}

Even though the runtime scaling of stochastic series expansion is linear in system size and inverse temperature, this example takes more computational effort than the previous one for the Ising model. It is therefore recommended to run it with MPI.
\begin{lstlisting}
$ mpirun -n $ncores julia example_job.jl run
\end{lstlisting}
Nevertheless, it is still tractable within a few minutes on a workstation or strong laptop and should yield the result file \texttt{bani2v2o8.result.json}. Like in the Ising example, we can use \texttt{Carlo.ResultTools} to plot the observable \texttt{MagChi} which corresponds to the static uniform magnetic susceptibility \cref{fig:susceptibility}.

We observe the dip feature that is caused by the easy-plane anisotropy $D^z$ and is also observed in experiment~\cite{Klyushina2021}. However, this feature is not quite converged with system size at $L=20$ and larger systems should be simulated in practice.

This concludes the present example, which showcases the effectiveness of a state-of-the-art Monte Carlo code built on Carlo.jl. Since this code was ported from an earlier C++ version, it also gives us an opportunity to directly compare the performance between the two languages, which we will do in \cref{sec:benchmarks}.
\subsection{Parallel tempering and the parallel run mode}
\label{sec:parallel_tempering}
Parallel tempering~\cite{Earl2005} is a technique that allows speeding up autocorrelation times in an ensemble of simulations by introducing additional Monte Carlo moves that exchange configurations between different simulation parameters. A typical application of this are glassy models, which can be simulated ergodically at high temperatures but freeze at low temperatures. If we simulate a chain of increasing temperatures in parallel, the probability distributions of adjacent temperatures may have some overlap so that higher temperature configurations may be accepted as updates in lower temperature simulations and vice versa. This leads to a steady exchange of configurations up and down the chain of temperatures, which can decrease the overall autocorrelation time significantly.

Carlo supports parallel tempering through the \texttt{ParallelTemperingMC} wrapper, which is a meta-implementation of the \texttt{AbstractMC} interface that allows to run other implementations with parallel tempering. We will first illustrate its usage with an example that is of interest to readers planning to use parallel tempering. Afterwards, we sketch its implementation as an example for implementing new algorithms with the parallel run mode feature in Carlo.jl.
\subsubsection{Usage}
Parallel tempering can be used with any implementation of the \texttt{AbstractMC} interface as long as it implements two additional functions,
\begin{lstlisting}[language=julia]
Carlo.parallel_tempering_log_weight_ratio(
    mc::MC, parameter::Symbol, new_value
)

Carlo.parallel_tempering_change_parameter!(
    mc::MC, parameter::Symbol, new_value
)
\end{lstlisting}
Both of them receive the name of the \texttt{parameter} along which parallel tempering is performed. In a tempering update, this parameter may be changed to a \texttt{new\_{}value}. The first function returns the logarithmic weight change $\log[W(p')/W(p)]$ that would result from changing the \texttt{parameter} from $p$ to $p'$. The second function puts that same change into action. Both the Ising reference implementation and \texttt{StochasticSeriesExpansion.jl} implement this interface and can be used with \texttt{ParallelTemperingMC}.

Parallel tempering is useful in situations where the untempered autocorrelation time of a simulation is longer than the time it takes to exchange replicas in the tempering process. Such a situation is not easy to achieve for the SSE algorithm in standard models due to its efficient global loop updates, but can be useful in more involved situations~\cite{Torres2020}. To keep the presentation of this usage example transparent, we will not introduce an involved model here and instead return to the Ising reference implementation that only contains local updates and is therefore known to have long autocorrelation times around the critical point.

Using \texttt{ParallelTemperingMC} requires just a few changes to the job script of \cref{code:input},
\begin{lstlisting}[language=Julia]
# parallel_tempering.jl

using Carlo
using Carlo.JobTools
using Ising

tm = TaskMaker()

tm.sweeps = 20000
tm.thermalization = 2000
tm.binsize = 100

Ts = range(1.5, 4, 30)

# contract temperatures around Tc for better distribution overlap
Tc = 2.269
Ts += 0.5 .* (Tc .- Ts) ./ (0.6 .+ (Tc .- Ts) .^ 2)


tm.parallel_tempering = (
    mc = Ising.MC,
    parameter = :T,
    values = Ts,
    interval = 1
)

Ls = [32, 64]
for L in Ls
    tm.Lx = L
    tm.Ly = L

    # tm.T is set implicitly by ParallelTemperingMC

    task(tm)
end

job = JobInfo(
    splitext(@__FILE__)[1],
    ParallelTemperingMC; # underlying MC set in tm.parallel_tempering
    run_time = "24:00:00",
    checkpoint_time = "30:00",
    tasks = make_tasks(tm),
    ranks_per_run = length(tm.parallel_tempering.values), # needs to match!
)

start(job, ARGS)
\end{lstlisting}
Most notably, we replace the MC algorithm in \texttt{JobInfo} by \texttt{ParallelTemperingMC} and use the parameter \texttt{tm.parallel\_{}tempering} to specify the details of how parallel tempering should be performed as follows
\begin{description}
\item[\texttt{mc}] the underlying model MC implementation.
\item[\texttt{parameter}] name of the parameter along which parallel tempering is performed, e.g. \texttt{:T} for the temperature.
\item[\texttt{values}] a chain of values of the parameter, which is simulated in parallel with nearest neighbor replica exchange.
\item[\texttt{interval}] the number of sweeps between proposing replica exchanges.
\end{description}
Lastly, \texttt{ParallelTemperingMC} has to be run in parallel run mode, which as of Carlo v0.2.3 has to be enabled manually by setting the \texttt{ranks\_{}per\_{}run} option in \texttt{JobInfo}. In the case of \texttt{ParallelTemperingMC}, the number of ranks per run has to be set to the number of parameters in the parallel tempering chain.

Running the job script works just as in the previous examples,
\begin{lstlisting}
$ mpirun -n $N julia parallel_tempering.jl run
\end{lstlisting}
except running with MPI is required and the number of MPI ranks $N$ has to be equal to \texttt{m * ranks\_{}per\_{}run + 1} with some integer $m$. For the example above, $N=61$ will simulate the two length-30 parallel tempering chains for $L=32$ and $L=64$ in parallel.

\begin{figure}
\begin{center}
\includegraphics{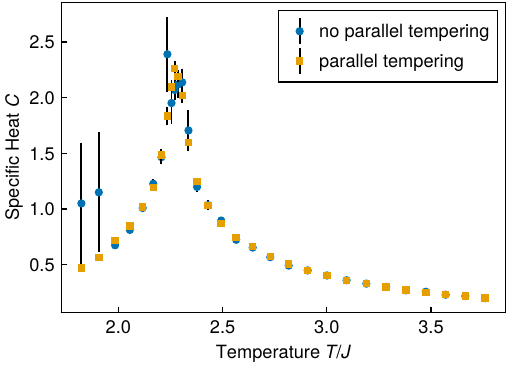}
\end{center}
\caption{\textbf{Parallel tempering in the Ising model.} The specific heat of the $64 \times 64$ square lattice Ising model as a function of the temperature, simulated with and without parallel tempering. Both simulations are perfored with 20000 measurement sweeps and 2000 thermalization sweeps.}
\label{fig:parallel_tempering}
\end{figure}

As shown in \cref{fig:parallel_tempering}, in this case, parallel tempering greatly decreases the statistical uncertainties close to the critical point and cures thermalization issues that arise at low temperatures. It should be mentioned that in the case of the Ising model, the critical slowing down can be more efficiently cured by introducing cluster updates~\cite{Wolff1989}. However, designing such global updates is not always possible for more complicated models.

\subsubsection{Implementation}
In the last subsection, we explained the usage of the parallel tempering feature. Now, we will now sketch its underlying implementation as an example of writing an algorithm with the parallel run mode feature of Carlo.jl.

In general, introducing parallel tempering into a simulation greatly increases the complexity of the parallel scheduling, since multiple simulations need to be orchestrated to run in lock step. However, Carlo.jl already supports this kind of parallelism through its parallel run mode. Parallel tempering (and similar meta-algorithms) can therefore be implemented as a thin layer on top of parallel run mode. 

To use parallel run mode, algorithms have to implement an alternative version of the \texttt{AbstractMC} interface, which essentially receives an additional MPI communicator (\cref{tab:parallelinterface}) that spans a team of MPI ranks that are collaborating on a single Carlo.jl run (that is, a single bar in \cref{fig:mcsched}).
\begin{table}
\begin{tabular}{l}
\hline
\textbf{Single run interface}\\
\hline
\texttt{MC(parameters::AbstractDict)}  \\
\texttt{init!(::MC, ::MCContext, params::AbstractDict)}\\
\texttt{sweep!(::MC, ::MCContext)}\\
\texttt{measure!(::MC, ::MCContext)}\\
\texttt{register\_{}evaluables(::Type{MC}, ::AbstractEvaluator, params::AbstractDict)}\\
\texttt{write\_{}checkpoint(::MC, ::HDF5.Group)}\\
\texttt{read\_{}checkpoint!(::MC, ::HDF5.Group)}\bigskip\\
\hline
\textbf{Parallel run interface}\\
\hline
\texttt{MC(parameters::AbstractDict)}$^*$\\
\texttt{init!(::MC, ::MCContext, params::AbstractDict, comm::MPI.Comm)}\\
\texttt{sweep!(::MC, ::MCContext, comm::MPI.Comm)}\\
\texttt{measure!(::MC, ::MCContext, comm::MPI.Comm)}$^{\dag}$\\
\texttt{register\_{}evaluables(::Type\{MC\}, ::AbstractEvaluator, params::AbstractDict)}\\
\texttt{write\_{}checkpoint(::MC, ::Union\{HDF5.Group,Nothing\}, comm::MPI.Comm)}$^\dag$\\
\texttt{read\_{}checkpoint!(::MC, ::Union\{HDF5.Group,Nothing\}, comm::MPI.Comm)}$^\dag$\\
\end{tabular}
\caption{Comparison between the standard single run interface for a type \texttt{MC <: AbstractMC}, where one Carlo.jl \textit{run} (i.e. one simulation of a single parameter set) is handled by one MPI rank, to the parallel run interface, where the work of one run is shared among a team of MPI ranks. The communicator \texttt{comm} passed as an additional argument spans this team of ranks. $^*$The constructor has the same signature, but additionally, \texttt{params[:\_{}comm]} contains the MPI communicator. $^\dag$In these methods, only the 0th rank of \texttt{comm} is allowed to call \texttt{measure!(ctx,...)}, read, or write to HDF5. The second argument of the checkpointing functions is \texttt{nothing} on the other ranks.}
\label{tab:parallelinterface}
\end{table}
In the parallel tempering scenario, we thus assign one chain of the tempered parameter to one Carlo run, and write a meta-implementation of the parallel run \texttt{AbstractMC} interface that runs the simulations for each value of the parameter.

\begin{lstlisting}[language=julia]
mutable struct ParallelTemperingMC{T} <: AbstractMC
    parameter_name::Symbol
    parameter_values::Vector{T}
    tempering_interval::Int

    parallel_measure::ParallelMeasurements
    chain_idx::Int
    child_mc::AbstractMC
end
\end{lstlisting}
The name of the tempered parameter is stored in \texttt{parameter\_{}name} with the values stored in \texttt{parameter\_{}values}. Tempering updates are performed every \texttt{tempering\_{}interval} sweeps, and \texttt{ParallelMeasurements} is a helper object for performing distributed measurements. Each instance of \texttt{ParallelTemperingMC} corresponds to one link in the parallel tempering chain and a single parameter value. \texttt{chain\_{}idx} stores the current position of the link and \texttt{child\_{}mc} stores a child Monte Carlo implementation that does the simulation between the tempering updates.

Most of the implementation work is then to implement the interface functions and calling the respective function of the child implementation, making it believe that it is running normally. In the constructor and the initialization function, we retrieve the configuration and inject the current value of the tempering parameter into the normal parameter set before passing it on tho the child.
\begin{lstlisting}[language=julia]
function ParallelTemperingMC(params::AbstractDict)
    config = params[:parallel_tempering]
    MC = config.mc
    tempering_interval = config.interval

    comm = params[:_comm]
    chain_idx = MPI.Comm_rank(comm) + 1

    modified_params = deepcopy(params)
    modified_params[config.parameter] = config.values[chain_idx]

    return ParallelTemperingMC(
        config.parameter,
        collect(config.values),
        config.interval,
        ParallelMeasurements(),
        chain_idx,
        MC(modified_params),
    )
end

function Carlo.init!(
    mc::ParallelTemperingMC,
    ctx::MCContext,
    params::AbstractDict,
    comm::MPI.Comm,
)
    modified_params = deepcopy(params)
    modified_params[mc.parameter_name] = mc.parameter_values[mc.chain_idx]

    Carlo.init!(mc.child_mc, ctx, modified_params)
end
\end{lstlisting}
In the \texttt{sweep!} function we let the child do sweeps, performing a tempering update periodically.
\begin{lstlisting}[language=julia]
function Carlo.sweep!(
    mc::ParallelTemperingMC, ctx::MCContext, comm::MPI.Comm
)
    Carlo.sweep!(mc.child_mc,
        make_parallel_context(ctx, mc.parallel_measure))

    if ctx.sweeps % mc.tempering_interval == 0
        synchronize_measurements!(
            ctx, mc.parallel_measure, mc.chain_idx, comm
        )
        tempering_update!(mc, ctx, comm)
    end
    return nothing
end
\end{lstlisting}
The function \verb+make_parallel_context+ creates a modified \texttt{MCContext} that reroutes possible measurements using \verb+mc.parallel_measure+.

Then, periodically,
\verb+synchronize_measurements!+ performs the necessary communications that collects these measurements from different ranks and combines them into vectors in the correct order. This is necessary because in parallel run mode, only the zeroth rank of the \texttt{comm} is allowed to do measurements.
Afterwards, \texttt{tempering\_{}update!} implements a standard neighbor swapping update~\cite{Earl2005} calling the functions \verb+parallel_tempering_log_weight_ratio+ to compare weight ratios and, if accepted, \verb+parallel_tempering_change_parameter!+ to switch the tempering parameter values. For further details, the interested reader is referred to the source code (\texttt{src/parallel\_{}tempering.jl}).

For the measurements, analogously, we call the \texttt{Carlo.measure!} function on the child.
\begin{lstlisting}[language=julia]
function Carlo.measure!(
    mc::ParallelTemperingMC, ctx::MCContext, comm::MPI.Comm
)
    Carlo.measure!(mc.child_mc,
        make_parallel_context(ctx, mc.parallel_measure)
    )
end
\end{lstlisting}
The \texttt{register\_{}evaluables} function works the same in single run and parallel run mode. There is no MPI communication as this function is called only once per run in the postprocessing phase, after the actual simulation is already done. For the parallel tempering implementation, this still takes some bookkeeping, as we need to string together the evaluables from the child implementations in a similar way as the measurements. We will skip these details here for the sake of brevity.

Finally, the checkpointing functions require some more MPI communication. The second argument in both of them is \texttt{nothing} on any rank except the zeroth rank of \texttt{comm}. We therefore communicate the state, given by the chain permutation, the parallel measurement object, and the internal state of the child implementations, before reading from or writing to the HDF5 file.
\begin{lstlisting}[language=julia]
function Carlo.write_checkpoint(
    mc::ParallelTemperingMC,
    out::Union{HDF5.Group,Nothing},
    comm::MPI.Comm,
)
    chain_permutation = MPI.Gather(mc.chain_idx, comm)
    child_mcs = MPI.gather(mc.child_mc, comm)
    parallel_measures = MPI.gather(mc.parallel_measure, comm)

    if MPI.Comm_rank(comm) == 0
        out["chain_permutation"] = chain_permutation

        for (i, (child_mc, parallel_measure)) in
            enumerate(zip(child_mcs, parallel_measures))
            Carlo.write_checkpoint(child_mc,
                create_group(out, "child_mcs/$i"))
            Carlo.write_checkpoint(parallel_measure,
                create_group(out, "parallel_measures/$i"))
        end
    end
    return nothing
end

function Carlo.read_checkpoint!(
    mc::ParallelTemperingMC,
    in::Union{HDF5.Group,Nothing},
    comm::MPI.Comm,
)
    child_mcs = MPI.gather(mc.child_mc, comm)

    if MPI.Comm_rank(comm) == 0
        chain_permutation = read(in, "chain_permutation")
        parallel_measures = [
            Carlo.read_checkpoint(ParallelMeasurements,
            in["parallel_measures/$i"])
            for i in eachindex(child_mcs)
        ]

        for (i, child_mc) in enumerate(child_mcs)
            Carlo.read_checkpoint!(child_mc, in["child_mcs/$i"])
        end
    else
        chain_permutation = nothing
    end

    mc.chain_idx = MPI.scatter(chain_permutation, comm)
    mc.child_mc = MPI.scatter(child_mcs, comm)

    mc.parallel_measure = MPI.scatter(parallel_measures, comm)

    parallel_tempering_change_parameter!(
        mc.child_mc,
        mc.parameter_name,
        mc.parameter_values[mc.chain_idx],
    )

    return nothing
end
\end{lstlisting}
This concludes the implementation of \texttt{ParallelTemperingMC} as an example of leveraging the parallel run mode feature. While some MPI communication and bookkeeping is required, the heavy lifting in terms of scheduling is being done behind the scenes by Carlo.jl. The main source of complexity in the parallel tempering implementation is the design requirement that it should work transparently with normal \texttt{AbstractMC} implementations. Utilities for this kind of interfacing, such as the \verb+ParallelMeasurements+ used above may also be useful for other meta-algorithms. While they are not currently part of the public Carlo.jl API, this may change in a future release.
\section{Benchmarks}
\label{sec:benchmarks}
\begin{figure}
\begin{center}
\includegraphics{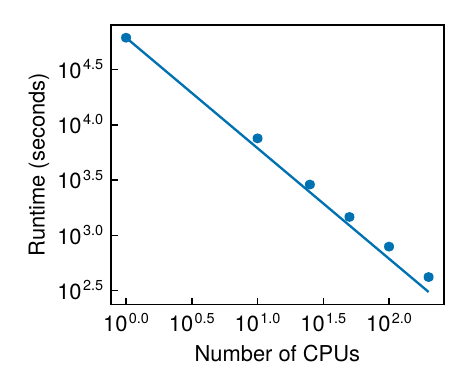}
\end{center}
\caption{\textbf{Runtime scaling with number of CPUs} of the example job script from \cref{code:input}, with the replacements \texttt{Ls = [100, 200, 300, 500]} and \texttt{sweeps = 200000}. The solid line shows the ideal linear speedup.}
\label{fig:benchmark}
\end{figure}
\begin{figure}
\begin{center}
\includegraphics{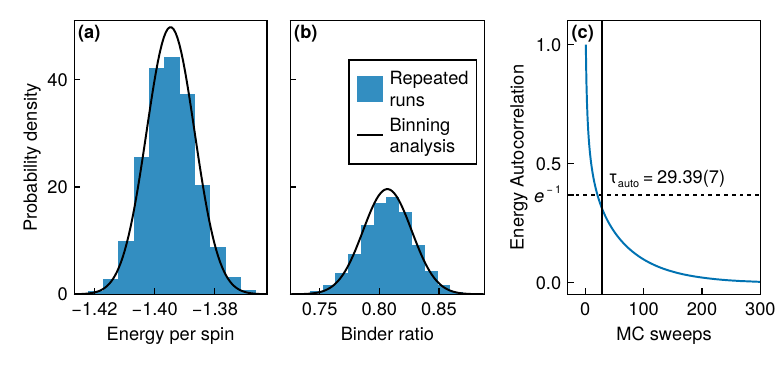}
\end{center}
\caption{\textbf{Binning analysis benchmark} using the $20\times 20$ square-lattice Ising model at temperature $T=2.3$, close to the critical point. The same simulation with 20000 sweeps and 4000 thermalization sweeps is repeated 9000 times. We then compare the distribution of these results to the distribution predicted by the binning analysis. (a) Histogram of the energy per spin compared to a Gaussian distribution with the average mean and standard deviation predicted from the binning analysis. (b) Histogram of the Binder ratio compared to a Gaussian distribution with the average mean and standard deviation predicted from jackknifing. (c) The energy autocorrelation function, averaged over all runs, compared to the autocorrelation time $\tau_\text{auto}$ predicted by Carlo.jl.}
\label{fig:statistics}
\end{figure}
In this section, we provide benchmarks to show (i) the runtime scaling of the parallel scheduler and (ii) the validity of the statistical postprocessing in the face of strongly autocorrelated data. For both, we use the Ising example code implemented in \cref{sec:ising}. Finally, we (iii) directly compare the performance of the Julia and the C++ versions of the stochastic series expansion code.

\subsection{Scheduler runtime scaling}
For the runtime scaling, we consider a workload inspired by \cref{code:input}, where we make the replacements \texttt{Ls = [100, 200, 300, 500]} and \texttt{sweeps = 200000} to increase the computational effort. Since the Ising code scales quadratically with the linear system size \texttt{Lx}, the tasks of this workload are quite imbalanced, comparable to the situation in \cref{fig:mcsched}. Nevertheless, we observe nearly linear scaling of the runtime with number CPUs used~(\cref{fig:benchmark}). We expect that the deviation from perfect scaling in this case is dominated by the serial cost of thermalization, which, based on the number of sweeps, is about 1\% of the total serial runtime.

\subsection{Validity of the postprocessing}
To confirm the accuracy of the statistical postprocessing, we consider the Ising model with $L=20$ and $T=2.3$, close to the critical point. In this region, the local updates of the Metropolis algorithm that we implemented are quite inefficient, leading to long autocorrelation times that require a careful analysis to retrieve the correct statistics.

To obtain the statistics of this model, there are two ways. The first one is to run many identical, independent simulations and create a histogram of the results. The second, more practical approach is to run a single simulation and perform a binning analysis on the correlated time series. By comparing the histogram of many independent simulations with the average results from the binning analysis implemented in Carlo.jl (\cref{fig:statistics}(a)), we confirm that the binning procedure outlined in \cref{sec:postproc} yields the correct errors for observables. As an example for the jackknifing feature, the Binder ratio (\cref{fig:statistics}(b)) also shows correct statistics. Finally, in \cref{fig:statistics}(c) we show the autocorrelation function for the energy per spin. We find that the autocorrelation time estimated from the binning analysis using~\cref{eq:autocorr} is a good estimate for the true decay of the autocorrelations.

\subsection{Performance comparison between Julia and C++}
The StochasticSeriesExpansion.jl package mentioned in \cref{sec:sse} is a Julia translation of the earlier C++ code~\cite{WeberFrust2024} that was used in Refs.~\cite{Jimenez2021,Weber2022,WeberCluster2022,WeberThermal2022,weber_cavityrenormalized_2023}. Since many of its inner workings are identical and both codes have seen a similar amount of optimization efforts, they provide an ideal opportunity for benchmarking the performance of Julia for quantum Monte Carlo applications, which we do in the following.

Although the codes behave identically at the core, a number of details need to be taken into account for a quantitative performance comparison. Firstly, the models implemented for use with them are not the same. For example, the anisotropic magnet from \cref{sec:sse} is not implemented in the C++ code. A model that is implemented for both is the spin-1/2 fully-frustrated bilayer~\cite{MullerHartmannExact2000,Alet2016} in the interlayer dimer basis, which we therefore use here (accessible through the \texttt{ClusterModel} in StochasticSeriesExpansion.jl). Another, more technical intricacy are the length of the SSE operator string and the number of worm updates. Both are parameters that are automatically tuned by the algorithm during its thermalization phase. Both codes use exactly the same controller logic for this, however the result is subject to noise. To eliminate this noise, we set both of them to the same values, predetermined by an earlier simulation. Lastly, the C++ version used a custom implementation of the mt19937 Mersenne Twister~\cite{Matsumoto1998}. Carlo.jl uses Julia’s \texttt{Random.Xoshiro} by default, which for this comparison, we replace by \texttt{Random.MersenneTwister}. All of these settings are implemented in the job script from \cref{lst:language_comparison}. The C++ code was compiled with GCC 12.2.0 with \texttt{-march=znver4 -ffast-math} and link-time optimization enabled. The Julia version was 1.11.1. The two simulations were carried out on AMD EPYC 9474F CPUs with 16 MPI ranks at a time.
\begin{figure}
\begin{center}
\includegraphics{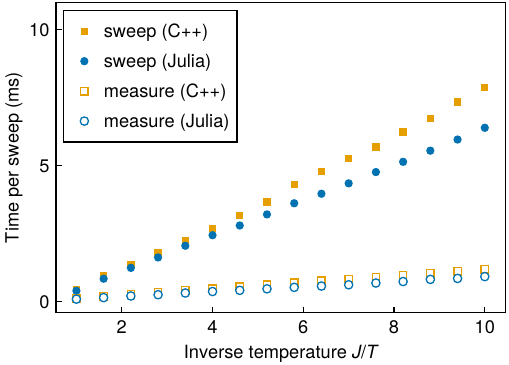}
\end{center}
\caption{\textbf{Language performance comparison} between the Julia and C++ implementations of the same SSE QMC algorithm. Shown are the average times spent on a sweep and a measurement as a function of inverse temperature $\beta=1/T$ for the fully frustrated bilayer with interlayer to intralayer coupling ratio $J/J_\perp = 2$ and $30\times 30$ dimer unit cells. }
\label{fig:language_comparison}
\end{figure}
\begin{lstfloat}
\begin{lstlisting}[language=julia]
using Carlo
using Carlo.JobTools
using StochasticSeriesExpansion
using Random

tm = TaskMaker()
tm.sweeps = 100000
tm.thermalization = 10000
tm.binsize = 100

tm.model = ClusterModel
tm.inner_model = MagnetModel
tm.cluster_bases = (StochasticSeriesExpansion.ClusterBases.dimer,)

tm.measure_quantum_numbers = [(;name=Symbol(), quantum_number=2)]
tm.parameter_map = (;
    J = vcat([:Jperp], repeat([:Jpar],8))
)

tm.Jpar = 1
tm.Jperp = 0.5

betas = range(1,10,16)
L = 30

tm.lattice = (
    unitcell = UnitCells.fully_frust_square_bilayer,
    size=(L,L)
)
for beta in betas
    tm.init_opstring_cutoff = 160000*beta/maximum(betas)
    tm.init_num_worms = 8
    tm.num_worms_attenuation_factor = 0.0
    tm.T = 1/beta
    tm.Lx = L
    tm.Ly = L
    task(tm)
end

job = JobInfo(
    "path/to/job",
    StochasticSeriesExpansion.MC;
    run_time = "24:00:00",
    checkpoint_time = "30:00",
    rng = Random.MersenneTwister,
    tasks = make_tasks(tm),
)

start(job, ARGS)
\end{lstlisting}
\caption{The StochasticSeriesExpansion.jl job file used for the performance comparison in \cref{fig:language_comparison}.}
\label{lst:language_comparison}
\end{lstfloat}

In \cref{fig:language_comparison}, both codes show the characteristic linear scaling of the SSE algorithm with inverse temperature. The Julia version shows slightly better performance than the C++ version for both sweeps and measurements. This should not be seen as proof that one language is inherently faster or slower, as further optimizations on either side are likely possible.

Some plausible explanations for the different performance could be the following. First, the sweep time could be partly affected by the different performance of the Mersenne Twister implementations. Second, due to just-in-time compilation, the Julia compiler has more compile-time knowledge. For example, both codes allow the SSE bond operators to connect an arbitrary number of sites. If the number of sites per operator is known at compile-time, which has been ensured in the Julia code, several expensive integer divisions can be replaced by cheaper bit shift operations. Achieving something similar in C++ is possible using sophisticated templating techniques, however, in the present code, this is only done in part. Lastly, the reason could be a random difference in the LLVM optimizer, used by Julia, and the GCC optimizer, used for C++ here, that happens to be relevant for the performance of the given code on the given architecture.

Despite these technical details that make performance comparisons of this kind hard to interpret, we can confidently conclude that it is possible to write state-of-the-art QMC codes in Julia without paying a price in performance.
\section{Conclusion}
\label{sec:conclusion}
Carlo.jl is a framework for writing Monte Carlo simulation codes in Julia. It features a Monte Carlo aware parallel scheduler with parallel tempering support, organized storage of input, checkpoint, and result files, as well as statistical postprocessing of the Monte Carlo results.

We have illustrated these features and the general usage of the framework by providing examples for the Ising model, the SSE QMC algorithm and parallel tempering. Our benchmarks show the efficient runtime scaling of the scheduler and the correctness of the statistical postprocessing. Additionally, since the SSE QMC code is based on an earlier C++ implementation, we performed a language performance comparison, in which the Julia version was faster, demonstrating the applicability of Julia for Monte Carlo workloads.

Based on these features and properties, the availability of a framework like Carlo.jl should enable the quick development of new user-friendly high-performance Monte Carlo codes in Julia.
\section*{Acknowledgements}
Carlo.jl is inspired by an earlier C++ code~\cite{loadleveller} written by the author, which in turn was inspired by a legacy code used in Stefan Wessel's group.
All plots except \cref{fig:binder} and \cref{fig:susceptibility} were created using Makie.jl~\cite{DanischKrumbiegel2021}.

\paragraph{Funding information}
L.W. acknowledges support by the Deutsche Forschungsgemeinschaft (DFG, German Research Foundation) through grant WE 7176-1-1.
The Flatiron Institute is a division of the Simons Foundation.
\begin{appendix}
\numberwithin{equation}{section}
\end{appendix}

\bibliography{bibliography.bib}

\end{document}